\documentclass[12pt]{iopart}
\usepackage{epsfig}
\usepackage{graphicx}
\usepackage{amsfonts}
\usepackage{amssymb}
\usepackage{graphicx}
 \usepackage{wrapfig}
\usepackage{textcomp}
\usepackage{subcaption}
\usepackage{hyperref}
\usepackage{pgf,pgfarrows,pgfnodes,pgfautomata,pgfheaps,pgfshade}
\usepackage[english]{babel}
\usepackage[english]{babel}
\newcommand{{\sign}}{\rm sign}
\newcommand{{\const}}{\rm Const}

\newcommand{\diag}{\rm diag}

\begin{document}
\title[]{Dispersive Response of a Disordered Superconducting Quantum Metamaterial}
\author{Dmitriy S. Shapiro$^{1,2,3}$, Pascal Macha$^{4,5}$, Alexey N. Rubtsov$^{1,3}$
 and  Alexey V. Ustinov$^{1,4,6}$}
\address{$^1$ Russian Quantum Center,  Novaya St.  100, Skolkovo, Moscow region, 143025, Russia}
\address{$^2$ Kotel'nikov Institute of Radio Engineering and Electronics of the Russian Academie of Sciences, Mokhovaya 11/7, Moscow,  125009, Russia}
\address{$^3$ Center of Fundamental and Applied Research, N.L. Dukhov All-Russia Institute of Science and Research, Sushchevskaya 22, Moscow, 123055, Russia\\
$^4$ Physikalisches Institut, Karlsruhe Institute of Technology, D-76128 Karlsruhe, Germany\\
$^5$ ARC Centre for Engineered Quantum Systems, University of Queensland, Brisbane, Queensland 4072, Australia\\
$^6$ National University of Science and Technology MISIS, Leninsky prosp. 4, Moscow, 119049, Russia}

\ead{shapiro@cplire.ru}
\begin{abstract}
We consider a disordered quantum metamaterial
formed by an array of superconducting flux qubits coupled to microwave photons in a cavity. We map the system on the Tavis-Cummings model accounting for the disorder in frequencies of the qubits. The complex transmittance is calculated with the parameters taken from state-of-the-art experiments.
We demonstrate that photon phase shift measurements allow to
distinguish individual resonances in the metamaterial with up to
100 qubits, in spite of the decoherence spectral width being remarkably larger than the effective coupling constant. Our simulations are in agreement with the results of the recently reported experiment \cite{Macha}.
\end{abstract}


\maketitle

\section{Introduction}

The novel type of quantum metamaterials  \cite{Rakhmanov, Fistul} composed of arrays of superconducting two-level systems (qubits) offers a platform for quantum simulators  and quantum memories \cite{DiCarlo,Nation}. Quantum metamaterials can be employed as test bench for studies of fundamental phenomena as ensemble  quantum  electrodynamics and spin resonance physics on macroscopic level \cite{Blais,YouNori}.  The flux qubit \cite{Orlando,mooij}, which can be viewed as a artificial atom \cite{Astafiev}, is a $\mu$m-sized superconducting loop with several Josephson junctions acting as nonlinear circuit elements. The strongly anharmonic potential of the flux qubit results in an effective two-level structure of the lowest pair of energy levels of the system, where the excitation frequencies fall into GHz range. Ground and excited levels of the qubit correspond to quantum superpositions of states with opposite directions of macroscopic persistent currents in the qubit loop. When coupled to a photon field in a superconducting cavity, the qubit becomes "dressed" with photons. The experimental studies of superconducting qubit-cavity quantum electrodynamics have shown vacuum Rabi oscillations \cite{Vijay, Fink}, spin echo and Ramsey fringes \cite{Bertet, Wallraff}, emission of single microwave photons \cite{Romero}, and Lamb shift \cite{Lamb}.

While the artificially made ensembles of superconducting qubits have an unavoidable spread of excitation frequencies, in contrast to identical natural atoms, they are easily tunable by varying an external magnetic flux. This fact makes possible observation of the  fundamental phenomena as dynamical Casimir effect by applying of a non-stationary external drive \cite{DCE1, DCE2}, under which GHz photons are created from cavity vacuum fluctuations.   In case of the multi-qubit system, a cm-long  cavity is used  as a transmission line where the virtual photon exchange provides a long-range qubit-qubit interaction in a sub-wavelength metamaterial. It was shown by Tavis and Cummings \cite{TavisCummings} that  $N$ identical two-level systems coupled to the single photon mode generate a  collective enhancement of the coupling constant proportional to $\sqrt{N}$. In a qubit array with disordered values of excitation frequencies photons are also coupled to a collective superradiant mode and decoupled from other $N-1$ subradiant modes. Collective enhancement combined with a possibility of tuning the parameters of the metamaterial by external magnetic field opens new possibilities for quantum information technology. For example, it has been suggested, that by applying gradients of magnetic field along the array of qubits, the system can be operated as a quantum memory with information encoded in  collective qubit states \cite{Moelmer}.

The present work is motivated by recent experiments by Macha et al. \cite{Macha} where the collective states of an array of superconducting flux qubits were probed by microwave photons transmitted through the resonator, in the regime of relatively strong decoherence and disorder in qubit excitation frequencies. The experiment has  shown that resonant photon phase shift effect is quite prominent and stable against the decoherence.  We perform simulations of the phase shift as a function of external flux that modifies the qubit excitation energies, in the setting similar to Ref. \cite{Macha}. In our approach, we perform a diagonalization of the Hamiltonian  in a single excitation basis. We  analyze the phase shift portrait for  different number of qubits and various values of disorder in qubit excitation frequencies. We show that photon phase shift measurements are capable to resolve individual resonances for up to $N=100$ qubits. Our simulations closely reproduce the experimental observations.

\section{Model}

\subsection{Hamiltonian of Disordered  Metamaterial}

\begin{figure}[h]
\begin{minipage}{\linewidth}
\center\includegraphics[height=0.4\linewidth]{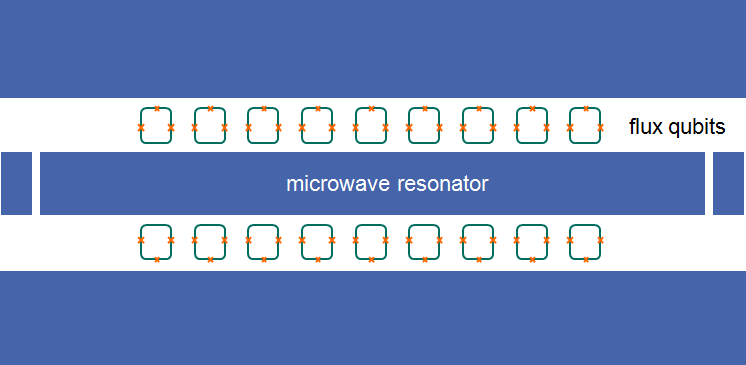}
\caption{ \label{fig0} Sketch of quantum metamaterial formed by an array of flux qubits embedded into a microwave resonator.}
\end{minipage}
\end{figure}

Considering the interaction of $N$  qubits  with a photon in a resonator (see Fig. \ref{fig0}), we start from the  Tavis-Cummings  Hamiltonian, which includes qubit-qubit interaction term
\begin{equation}
H=\omega_ra^+a+\sum\limits_{i=1}^N\epsilon_i\sigma^+_i\sigma^-_i+\sum\limits_{i=1}^Ng_i(\sigma^+_i a + a^+\sigma^-_i)+\sum\limits_{i=1}^{N-1}g_{qq,i}\sigma^+_{i+1}\sigma^-_i \label{h0}
\end{equation}
Bosonic operators $a^+, a$  in (\ref{h0}) correspond to the photon mode $\omega_r$ in the cavity and the single $i$-th qubit is  described  in terms of  uppering and lowering Pauli operators $\sigma_i^+=|e\rangle\langle g|, \sigma_i^-=|g\rangle\langle e|$ acting on its ground and excited states.
The first and the second terms  in Eq. (\ref{h0}) correspond to photon and qubit excitations of frequencies $\omega_r$ and $\epsilon_i$, the third one is qubit-cavity coupling written in rotating wave approximation.  The last term is an effective direct nearest neighbor qubit-qubit interaction.

The  qubit excitation frequency  $\epsilon$ follows from  microscopic  Josephson and charging energies, $E_J$ and $ E_C$, which imply  $E_J> E_C$ in the operating  regime of flux qubit. When the external magnetic flux  threading the qubit loop is equal to half of the magnetic flux quantum $\Phi=\Phi_0/2$, with $ \Phi_0=h/(2e)$,  the total energy of the qubit has two symmetric minima related to the opposite  circulations of persistent currents $|{\downarrow}\rangle$ and $|{\uparrow}\rangle$ where the  tunneling rate $\Delta$ between two wells  depends on $E_J$ and $E_C$. The condition that $\Delta$ is higher than dephasing rate $\Gamma_\varphi$ allows for quantum superpositions of $|{\downarrow}\rangle$ and $|{\uparrow}\rangle$ resulting in two non-degenerate states $|g\rangle,|e\rangle$ being the basis of (\ref{h0}). By detuning the magnetic flux from $\Phi=\Phi_0/2$ one shifts the two minima by the energy
\begin{equation}
 \varepsilon_i(\Phi)=\frac{2I_{p,i}}{\hbar}\left(\Phi-\frac{\Phi_0}{2}\right),                                                                                                                                                                                                                                                                                                                                                                                                                                                                                                                                                                                                                                                                                                                                                                                                                                                                                                                                                                                                                                                                                                                                                                                                                                                                                                                                                                                                                                                                                                                                                                                                                                                                                                                                                                                                                                                                                                                                                                                                                                                                                                                                                                                                                                                                                                                                          \end{equation}
 where $I_{p,i}$ is the $i$-th qubit nominal current \cite{Orlando}. The excitation frequency of $i$-th qubit is given by the Floquet relation
\begin{equation}
\epsilon_i(\Phi) = \sqrt{\Delta_i^2+\varepsilon_i(\Phi)^2}.
\end{equation}
The qubit-cavity coupling constants in Eq. (\ref{h0}) are renormalized bare constants $g_i=\frac{\Delta_i}{\epsilon_i(\Phi)}g_i^{bare}$ written in the rotated  basis $|g\rangle,|e\rangle$. Everywhere below we work with the effective constants.
Under uniform flux biasing conditions, the  spread in $\epsilon_i$ depends mainly on qubit excitation gaps $\Delta_i$, rather than $I_{p,i}$, due to  exponential dependence of $\Delta_i\propto \sqrt{E_J/E_C} \exp(-\alpha\sqrt{E_J/E_C})$ on the $E_J/E_C$ ratio which fluctuates from one flux qubit to another.  The system under consideration \cite{Macha} is strongly disordered because the spread of excitation gaps $\Delta_i$ is estimated as high as $\sigma_{\Delta }\leq 20\%$. The system has  an intermediate collective coupling  strength value $\Omega$, which is  smaller than decoherence rate $\Gamma_\varphi$ but larger than relaxation rate $\kappa$ of the cavity
$$
\Gamma_\varphi > \Omega>\kappa, \quad \Omega=\sqrt{\sum\limits_{i=1...N} g_i^2}.
$$
In the experiment by Macha et al. \cite{Macha}, the external magnetic flux $\Phi$ tunes qubit excitation frequencies in resonance with the cavity mode  providing the shift of the  phase $\varphi(\Phi)$  of a weak external probe signal at  cavity mode frequency $\omega_r/2\pi=7.78$ GHz.  From the measured $\varphi(\Phi)$ it was found that number of qubits in an ensemble, which form a collective mode, corresponds to $N=8$. Relevant parameters of the studied metamaterial are the following: qubit-cavity coupling strength $g_i/2\pi=(1.2\pm 0.1)$ MHz, dephasing $\Gamma_\varphi/2\pi=55$ MHz, persistent current $I_p=(74\pm 1)$ nA and average value of the excitation gap  $\Delta/2\pi=5.6$ GHz.
The qubit-qubit interaction, estimated as $g_{qq,i}/2\pi<1$ MHz, has negligible effect  on $\varphi(\Phi)$. The interaction only slightly shifts energies of collective states and does not contribute directly to the collective qubit-cavity coupling strength $\Omega$, which is the most relevant for the phase shift.

\subsection{Exact Diagonalization Approach}

 The qubit dephasing in the system under consideration leads to smearing of the photon density of states which becomes not informative. However, the experimentally measured phase locking effect is quite prominent  \cite{Macha}. In our work here, we focus on the influence of relatively large diagonal disorder  in the qubit excitation frequencies and their number $N$ on the phase shift  $\varphi(\Phi)$ of the transmitted  photons. We calculate the phase shift from a complex phase of the photon Green function  $\varphi(\Phi)=\arg \mathcal{D}_\omega$, where  $\omega$ is the probe frequency  and  $\Phi$ is the external flux, counted from the symmetry point value $\Phi_0/2$, which shifts all excitation energies in the qubit ensemble.

Considering approximately resonant regime where the excitation frequencies of qubits and resonator mode are close to each other
\begin{equation}
  |\epsilon_i-\omega_r |< \omega_r \label{res1}
\end{equation}
we find a spectrum of the system in a regime of single excitation by means of exact diagonalization of the Hamiltonian. This regime is realized under experimental conditions due to small microwave probe power and low temperature of the system. Our solution corresponds to the fully quantum regime where all the variables $\sigma_i^-, a $ are  quantum fields.

Below we  introduce the basis of $N+1$  states of approximately equal energies which are related to single excitation either in the photon cavity or in one of  $N$ qubits. Namely, the state where a single photon is excited and all qubits are in the ground state reads
$$
| \mathbf{1} \rangle=\underbrace{
| 1; 0,   \ldots,  0 \rangle}_{N+1}
$$
The excitation in $1$-st, $i$-th or $N$-th qubits without the photon are given by
$$
| \mathbf{2} \rangle=| 0; \ 1, 0 ,  \ldots ,0 \rangle,
$$

$$
| \mathbf{i+1} \rangle=| 0; \ 0,   \ldots , 1,  \ldots ,0 \rangle
$$
and
$$
 | \mathbf{N+1} \rangle=| 0;  \ 0,  \ldots , 0, 1 \rangle .
$$
The Hamiltonian  (\ref{h0}) in a matrix form $\mathbf{H}_{i,j}=\langle\mathbf{i} | H | \mathbf{j} \rangle$ in  terms of this  single excitation basis $| \mathbf{ i } \rangle$ reads

$$
\mathbf{H}_{i,j}=
%\begin{pmatrix}
\left[\matrix{\omega_r & g_1 & g_2 &  g_3 & g_4 & g_5 & \ \ldots \cr \cr
g_1 & \epsilon_1 & g_{qq,1} & 0 & 0 & 0 & \ \ldots \cr \cr
g_2 & g_{qq,1}  & \epsilon_2 & g_{qq,2} & 0 & 0 & \ \ldots \cr \cr
g_3 & 0 & g_{qq,2}  & \epsilon_3 & g_{qq,3} & 0 &    \ \ldots \cr \cr
g_4 & 0 & 0 & g_{qq,3}  & \epsilon_4 & g_{qq,4} &   \ \ldots \cr \cr
g_5 & 0 & 0 & 0 & g_{qq,4} & \epsilon_5 &    \ \ldots \cr \cr
\vdots & \vdots & \vdots & \vdots & \vdots & \vdots  &  \ \ddots} \right].
%\end{pmatrix}.
$$
The photon Green function $\mathcal{D}_\omega$ is given by the first diagonal $\{ 1, 1\}$ element of the resolvent  matrix $\mathbf{G}$
\begin{equation}
\mathbf{G}_{i,j}=[\omega\delta_{i,j}-\mathbf{H}_{i,j}-{\diag}(i\kappa, i\Gamma_\varphi,...)]^{-1}.
\end{equation}
This is the matrix of retarded Green functions, where the last term ${\diag}(i\kappa, i\Gamma_\varphi,...)$ introduced in order to take into account the damping in the cavity  $\kappa/2\pi=715$ kHz and dephasing of qubits. We note that the above expression for the photon Green function  $\mathcal{D}_\omega$  and $\mathbf{G}_{i,j}$ allowing us to calculate the phase shift coincides with the generic equations obtained earlier by Volkov and Fistul \cite{Fistul}.

\section{Results}

\subsection{Strong Disorder and Large $N$}

In our numerical studies, we model the system with large amount of qubits and in the presence of disorder between them. Namely, we fix metamaterial parameters and plot the phase shift  increasing qubit number $N$, but keeping spread of  $\Delta_i$ constant, see Fig. \ref{fig2}, and vice versa, see Fig. \ref{fig3}.
The reported results for the $\varphi(\Phi)$ are obtained at the experimental values $\omega/2\pi=\omega_r/2\pi=7.78$ GHz, $I_p=74$ nA and the following parameters: decoherence rate $\Gamma_\varphi/2\pi=33$ MHz, average qubit excitation gaps  $\Delta/2\pi=5.9$ GHz with normal distribution and  dispersion $\sigma_\Delta=3.6 \%$, effective qubit-cavity coupling $g/2\pi=1.1$ MHz with small dispersion $\sigma_g=1\%$ and $N=7$ qubits in the ensemble. These parameters give best fit in the experimental regime at small $N$ discussed in the next section \ref{exp_res}.

Figure \ref{fig2} shows plots of the phase shift dependence on magnetic flux for $N=20, 50, 100, 250, 500$ qubits in the system and fixed spread $\sigma_\Delta=3.6 \%$ of the normal distribution of $\Delta_i$.
One can see that for $N<100$ the phase shift fluctuates and depends
 on a particular realization of the qubit frequencies, while for $N>100$  it appears rather smooth and regular, meaning that the array of qubits can be treated as a continuous system at these parameters.
In Fig. \ref{fig3}, we show results
for different values of disorder with the following dispersions $\sigma_\Delta=5, 7.5, 10, 15, 25 \%$, keeping  qubit number $N=250$ fixed. One can see from the Fig. \ref{fig3} that, with increasing disorder, the phase shift pictures start to reveal fluctuations. For the realistic spread $\sigma_\Delta=10\% $ the system remains far from the ensemble limit, even for this relatively large number of qubits.
\begin{figure}[h]
\begin{minipage}{\linewidth}
\center {\vspace{1cm}\includegraphics[height=0.4\linewidth]{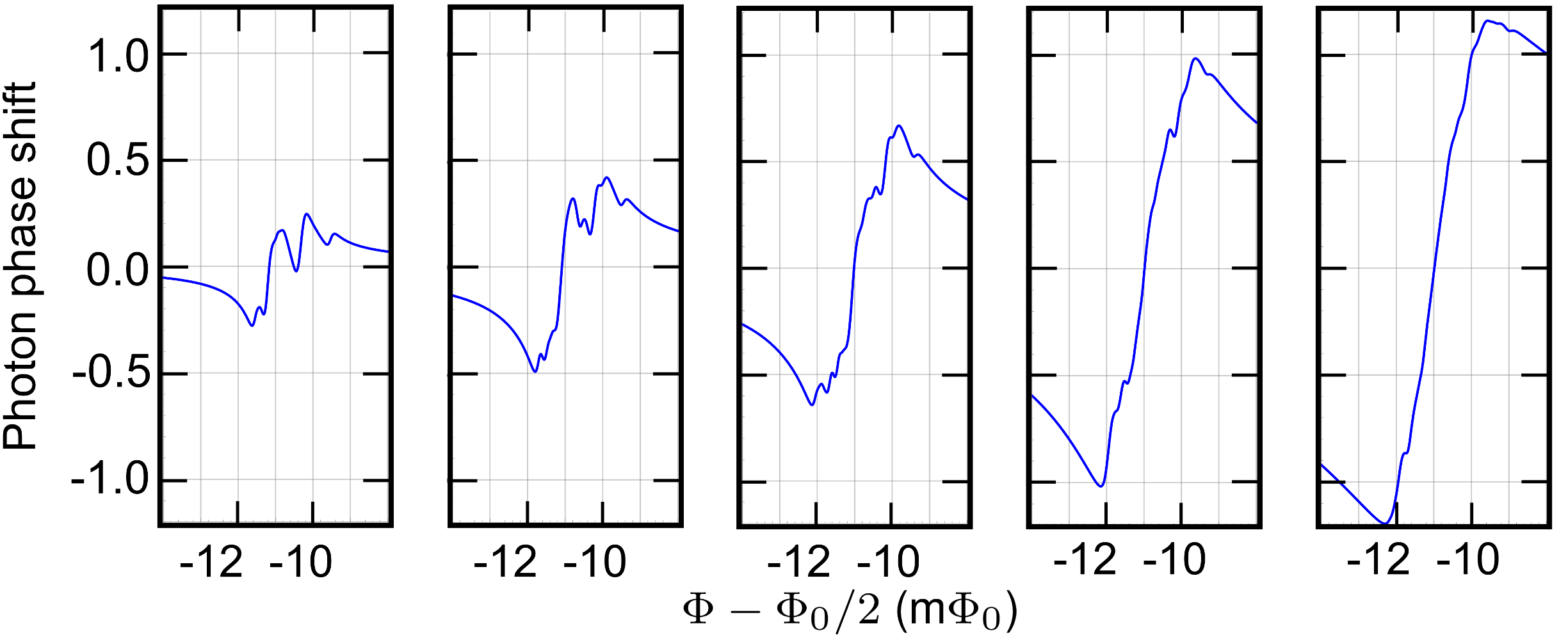} }
 \label{fig2}
\end{minipage}
\caption{ \label{fig2} Numerical results for the phase shift $\varphi(\Phi)$ at qubit energy gap disorder $\sigma_\Delta=3.6\%$ of $N=20, 50, 100, 250, 500$ qubits.}
\end{figure}
 \begin{figure}[h]
\center\includegraphics[height=0.4\linewidth]{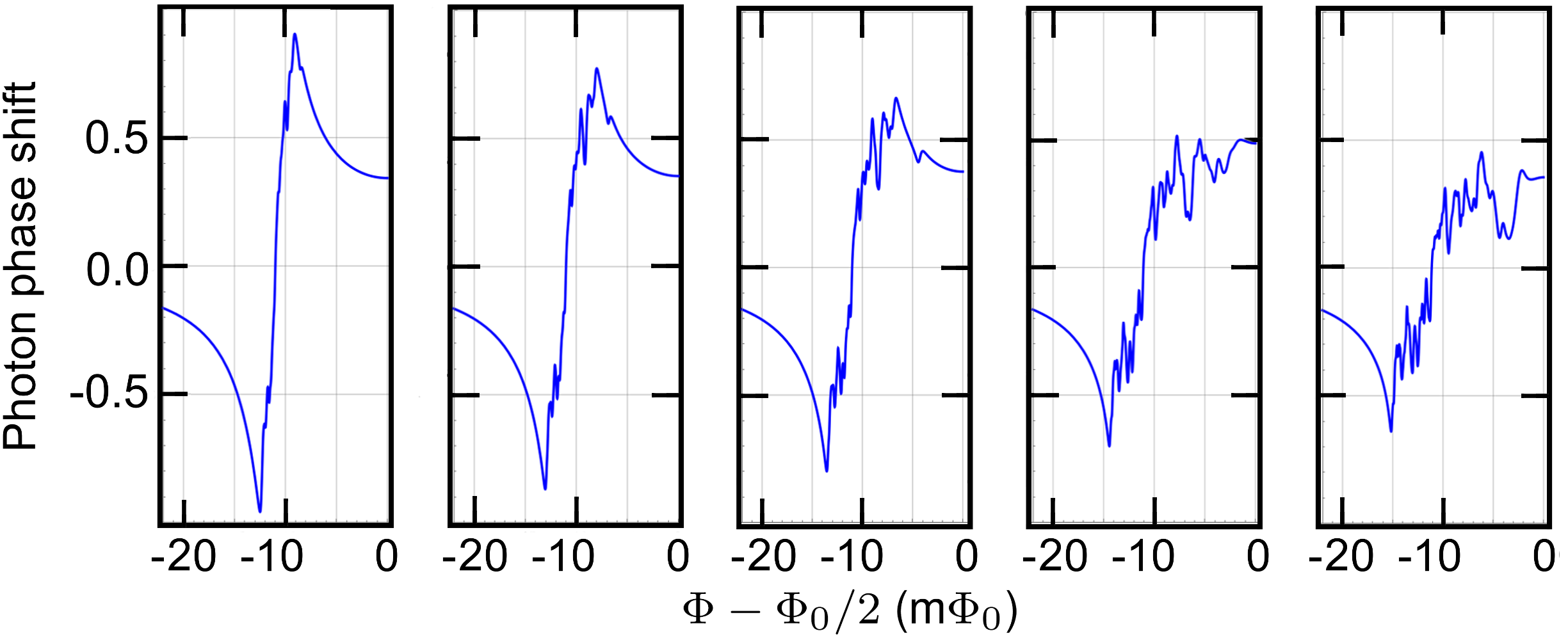}
\caption{ \label{fig3} Numerical results for the phase shift $\varphi(\Phi)$ of $N=250$ qubits using five different values of disorder  $\sigma_\Delta=5, 7.5, 10, 15, 25 \%$. Parameters of the system:  probe frequency  $\omega/2\pi=\omega_r/2\pi=7.78$ GHz and relaxation rate $\kappa/2\pi=715$ kHz, average excitation gap  $\Delta/2\pi=5.9$ GHz of qubits with normal distribution and dispersion $\sigma_\Delta=3.6\%$, average coupling constant $g/2\pi=1.1$ MHz with $\sigma_g=1\%$, decoherence $\Gamma_\varphi/2\pi= 33$ MHz, persistent current $I_p=74$ nA. }
\end{figure}

\label{num_res}

\subsection{Experimental Regime, Small $N$}

\begin{figure}[h]
\begin{minipage}{\linewidth}
\center\includegraphics[height=0.4\linewidth]{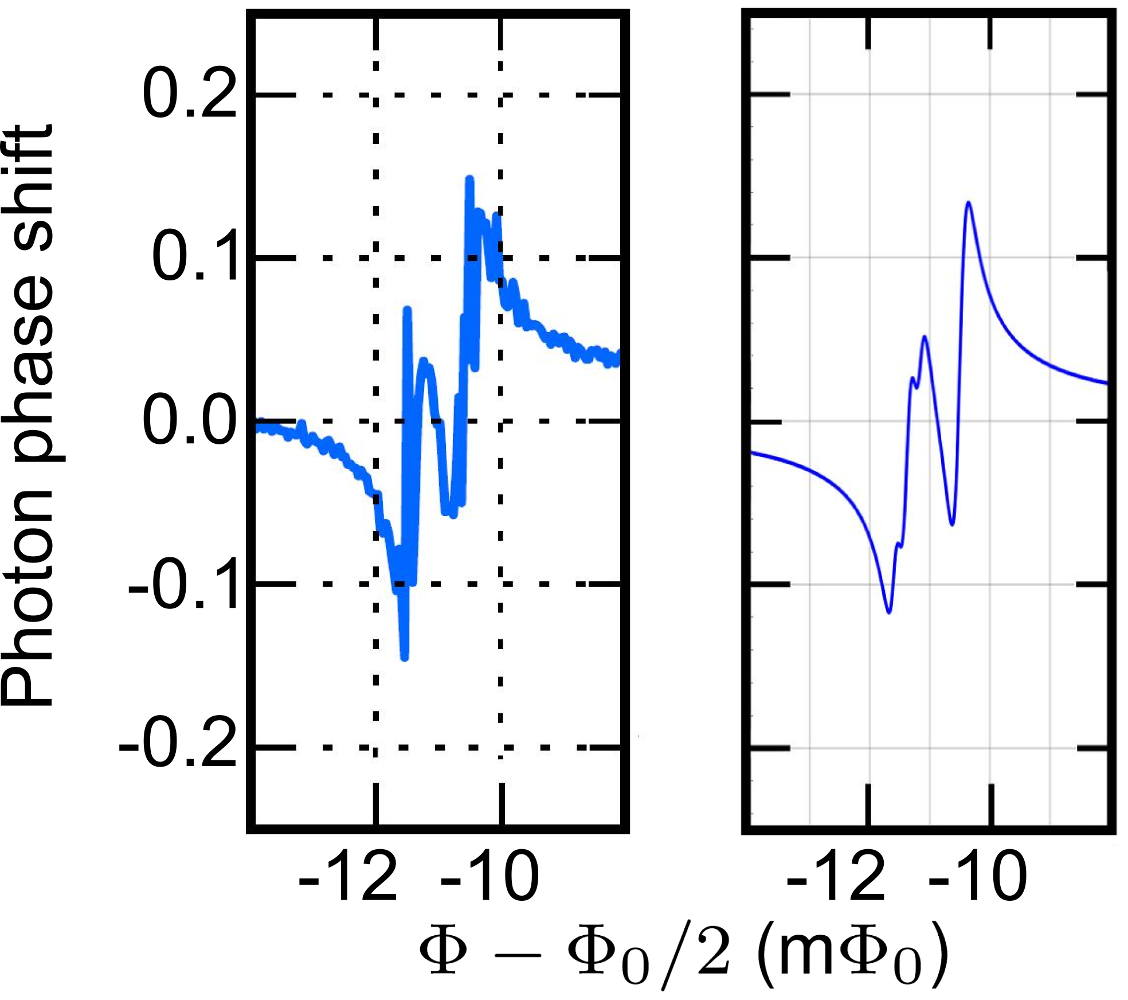}
\caption{ \label{fig1} Experimental (left panel) and numerical (right panel) results for the photon phase shift $\varphi(\Phi)$ as a function of external flux bias $\Phi$, calculated for the system of $N=7$ qubits.}
\end{minipage}
\end{figure}

Next, we compare experimental  \cite{Macha} (Fig. \ref{fig1}, left panel) and numerical (Fig. \ref{fig1}, right panel) results for the phase shift $\varphi(\Phi)$.
The resonator frequency $\omega/2\pi=\omega_r/2\pi=7.78$ GHz and persistent current $I_p=74$ nA are fixed. In our numerical method, we assume a normal distribution of the qubit excitation gaps and effective coupling. We selected a parameter distribution with dispersions $\sigma_\Delta=3.6 \%$ and $\sigma_g=1\%$ for a system containing  $N=7$ qubits with an average excitation gap $\Delta/2\pi=5.9$ GHz, which closely resembles the experimental data. Subsequently, we fitted the decoherence rate and effective qubit coupling and found $\Gamma_\varphi/2\pi=33$ MHz and $g/2\pi=1.3$ MHz, respectively.
While the assumed spread is less than the experimentally expected one ($\sigma_\Delta^{exp}<20 \%$) all found parameters are in correspondence with those quoted in \cite{Macha}, where $\Delta^{exp}/2\pi=5.6$ GHz, $g^{exp}/2\pi=1.2$ MHz, $\Gamma_\varphi^{exp}/2\pi=55$ MHz and $N^{exp}=8$. The values for qubit number and dephasing reported in Ref. \cite{Macha} were found under the assumption of identical qubits. Here we show, that the experimental data can be reproduced fairly well under the assumption of randomly distributed qubit parameters. In the experiment it was observed, that an ensemble of  $N^{exp}=8$ qubits, resonantly interacting with the cavity mode and monitored over long time, spontaneously dissolved into two sub-ensembles of $4$ qubits each, resulting into  two jumps in Fig. \ref{fig1} (left panel). In our exact diagonalization procedure we do not find the formation of sub-ensembles but arrive at correspondence with the experimental curves  if parameters of the metamaterial and disorder are close to the experimental values.

\label{exp_res}

\section{Conclusions}

We theoretically studied the model of a flux qubit array coupled to a cavity, with disorder in qubit excitation frequencies. The system under consideration contains a finite number of qubits and operates in the intermediate regime where disorder range and decoherence rate exceed the effective qubit-cavity coupling. We calculated the photon Green function through the exact diagonalization of the Hamiltonian in the single excitation regime, assuming low power of external microwave driving. We found that the resonant phase shift of a transmitted probe signal shows quantitative correspondence with the experimental data \cite{Macha}. Variations in phase shift characteristics at different values of the disorder and number of qubits in the system were presented.

\section{Acknowledgments}

This work was supported by the Ministry for Education and Science of Russian Federation under contract no. 11.G34.31.0062 and in the framework of Increase Competitiveness Program of the National University of Science and Technology MISIS under contract no. K2-2014-025.


\section*{References}
\begin{thebibliography}{999}
 \bibitem{Macha}  Macha, P.; Oelsner, G.;  Reiner, J.-M.; Marthaler, M.; Andr\'e, S.; Sch\"on, G.; H\"ubner, U.;  Meyer, H.-G.; Il'ichev, E.; Ustinov, A.V.  Implementation of a quantum metamaterial using
superconducting qubits. {\em Nat. Commun.} {\bf 2014}, {\em 5}, 5146.
\bibitem{Rakhmanov} Rakhmanov, A.L.;  Zagoskin, A.M.;  Savel'ev, S.;  Nori, F.  Quantum
metamaterials: electromagnetic waves in a Josephson qubit line. {\em   Phys. Rev. B} {\bf 2008}, {\em 77}, 144507.
\bibitem{Fistul} Volkov, P.; Fistul, M. V. Collective quantum coherent oscillations in a globally coupled array of qubits. {\em   Phys. Rev. B} {\bf 2014},    {\em 89}, 054507.
\bibitem{DiCarlo}  DiCarlo, L.;  Chow, J. M.;  Gambetta, J. M.;  Bishop, L.S.;  Johnson, B.R.;  Schuster, D.I.;  Majer, J.;  Blais, A.;  Frunzio, L.;  Girvin, S.M.;  Schoelkopf, R.J. Demonstration of two-qubit algorithms with a
superconducting quantum processor. {\em   Nature} {\bf  2009}, {\em 460}, 240.
\bibitem{Nation} Nation, P. D.;  Johansson, J.R.;  Blencowe, M.P.;  Nori, F. Stimulating uncertainty: Amplifying the quantum vacuum
with superconducting circuits. {\em   Rev. Mod. Phys.} {\bf  2012}, {\em 84}, 1.
 \bibitem{Blais} Blais, A.; Huang, R.-S.;  Wallraff, A.;  Girvin, S.M.;  Schoelkopf, R.J. Cavity quantum electrodynamics for superconducting electrical circuits: An architecture for quantum computation. {\em   Phys. Rev. A} {\bf  2004}, {\em 69}, 062320.
\bibitem{YouNori} You, J.Q.; Nori, F.  Atomic physics and quantum optics using
superconducting circuits. {\em   Nature} {\bf  2011}, {\em 474},  589.
\bibitem{Orlando} Orlando, T. P.;    Mooij, J.E.;  Tian, L.;   van der Wal, C.H.;  Levitov, L.S.;  Lloyd, S.;  Mazo, J.J. Superconducting persistent-current qubit. {\em   Phys. Rev. B} {\bf  1999}, {\em 60}, 15398.
\bibitem{mooij} Mooij, J. E.;  Orlando, T.P.;  Levitov, L.;  Tian, L.;  van der Wal, C.H.; Lloyd, S. Josephson persistent-current qubit. {\em   Science} {\bf  1999},{\em 285}, 1036.
\bibitem{Astafiev} Astafiev, O.;  Zagoskin, A.M.;  Abdumalikov Jr., A.A.;  Pashkin, Yu.A.;  Yamamoto, T.;  Inomata, K.;  Nakamura, Y.;  Tsai, J.S. Resonance fluorescence of a single artificial atom. {\em   Science} {\bf  2010}, {\em 327}, 840.
\bibitem{Vijay} Vijay, R.;  Macklin,	C.;  Slichter, D.H.;  Weber,	S.J.;  Murch,	K.W.;  Naik,	R.;  Korotkov, A.N.;  Siddiqi, I.  Stabilizing Rabi oscillations
in a superconducting qubit using quantum feedback. {\em   Nature} {\bf  2012}, {\em 490},  77.
\bibitem{Fink}  Fink, J. M.;  Bianchetti, R.;  Baur, M.;  G\"oppl,  M.;  Steffen, L.;  Filipp, S.;  Leek, P.J.;  Blais, A.;  Wallraff, A. Dressed Collective Qubit States and the Tavis-Cummings
Model in Circuit QED. {\em   Phys. Rev. Lett.} {\bf  2009}, {\em 103}, 083601.
\bibitem{Bertet} Bertet, P.;   Chiorescu, I.;  Burkard,  G.;  Semba,  K.;  Harmans, C.J.P.M.;  DiVincenzo, D.P.;  Mooij, J.E. Dephasing of a Superconducting Qubit Induced by Photon Noise. {\em   Phys. Rev. Lett.}  {\bf  2005}, {\em 95}, 257002.
\bibitem{Wallraff} Wallraff, A.;  Schuster, D.I.;  Blais, A.;  Frunzio, L.;  Majer, J.;  Devoret, M.H.;  Girvin, S.M.;  Schoelkopf, R.J. Approaching Unit Visibility of a Superconducting Qubit with Dispersive Readout. {\em   Phys. Rev. Lett.} {\bf  2005}, {\em 95}, 060501.
\bibitem{Romero} Romero, G.;  Garc\'ia-Ripoll, J.J.;   Solano, E.  Microwave
photon detector in circuit QED. {\em   Phys. Rev. Lett.} {\bf  2009}, {\em 102}, 173602.
 \bibitem{Lamb} Fragner, A.; G\"oppl, M.;  Fink, J.M.;  Baur, M.;  Bianchetti, R.;  Leek, P.J.; Blais, A.;  Wallraff, A.  Resolving Vacuum Fluctuations
in an Electrical Circuit by Measuring the Lamb Shift. {\em    Science} {\bf  2008},
{\em 322}, 1357.
 \bibitem{DCE1} L\"ahteenm\"aki,  P.;    Paraoanua,  G.S.;  Hasselb, J.;  Hakonena, P.J. Dynamical Casimir effect in a Josephson metamaterial. {\em   Proc. Natl. Acad. Sci.} {\bf  2011}, {\em 111}, 10485.
 \bibitem{DCE2}  Wilson, C.M.;  Johansson,	G.;  Pourkabirian,	A.;  Simoen,	M.;  Johansson,	J.R.;  Duty,	T.;  Nori	F.;  Delsing, P. Observation of the Dynamical Casimir Effect in a Superconducting Circuit. {\em   Nature} {\bf  2011}, {\em 479}, 376.
\bibitem{TavisCummings} Tavis, M.; Cummings, F. W. Exact solution for an
n-molecule-radiation-field hamiltonian. {\em  Phys. Rev.} {\bf  1968}, {\em 170},
379.
\bibitem{Moelmer} Julsgaard, B.; M\o lmer, K. Fundamental limitations in spin-ensemble quantum memories for cavity fields.  {\em Phys. Rev. A} {\bf 2013}, {\em 88}, 062324.
\end{thebibliography}
\end{document}